\documentclass[twocolumn,showpacs,amsmath,amssymb,prl]{revtex4}

\usepackage{graphicx}
\usepackage{dcolumn}
\usepackage{bm}

\begin{document}

\title{Comment on "Test of constancy of speed of light with rotating cryogenic optical resonators"}
\author{Michael E. Tobar$^1$}
\email{mike@physics.uwa.edu.au}
\author{Peter Wolf$^{2,3}$}
\author{Paul L. Stanwix$^1$}
\affiliation{\\ $^1$University of Western Australia, School of
Physics M013, 35 Stirling Hwy., Crawley 6009 WA, Australia\\
$^2$SYRTE, Observatoire de Paris, 61 Av. de l'Observatoire, 75014 Paris, France\\
$^3$Bureau International des Poids et Mesures, Pavillon de Breteuil,
92312 S\`evres Cedex, France}

\date{\today}

\begin{abstract}
A recent experiment by Antonini et. al. [Phys. Rev. A {\bf 71}, 050101R 2005], set new limits on Lorentz violating parameters in  the frame-work of the photon sector of the Standard Model Extension (SME), $\tilde{\kappa}_{e-}^{ZZ}$, and the Robertson-Mansouri-Sexl (RMS) framework, $\beta-\delta-1/2$. The experiment had significant systematic effects caused by the rotation of the apparatus which were only partly analysed and taken into account. We show that this is insufficient to put a bound on $\tilde{\kappa}_{e-}^{ZZ}$ and the bound on $\beta-\delta-1/2$ represents a five-fold improvement not a ten-fold improvement as claimed.
\end{abstract}

\maketitle


Non-rotating experiments that test Lorentz Invariance are not sensitive to the SME parameter $\tilde{\kappa}_{e-}^{ZZ}$ \cite{Lipa,Muller,WolfGRG,Wolf04,Kosto1,KM}. To determine this parameter one requires active rotation with a non-zero signal expected to occur at twice the rotation frequency, $2\omega_R$ \cite{Ant,prl,bchap}. Thus, it is important to control and minimize systematic signals at this harmonic. If systematics dominate over statistical uncertainties, care must be taken when analyzing the data as it is difficult to distinguish the systematic signal from an actual non-zero value of $\tilde{\kappa}_{e-}^{ZZ}$. One way to do this is to characterize the systematic and subtract it from the data, which can be a difficult process as the amplitude may not necessarily be stationary over the period of data collection. Nevertheless, if one is careful it is a valid process and Antonini et. al. \cite{Ant} did effectively account for part of the unknown systematics in their experiment by subtracting the frequency modulation of the resonators induced by tilt. However, they were still left with a statistically significant amplitude at $2\omega_R$ which lead to a positive signal for Lorentz violation of $\tilde{\kappa}_{e-}^{ZZ} = (-2.0\pm0.2\times10^{-14})$. They state that this is likely due to a (non-accounted) systematic effect and thus claim an upper bound of $|\tilde{\kappa}_{e-}^{ZZ}|<2\times 10^{-14}$. 

Since the suspected systematic is uncharacterized there is no way to know if the measured amplitude is due to a systematic or a true non-zero value of $\tilde{\kappa}_{e-}^{ZZ}$. Furthermore, one can not rule out that the uncharacterized systematic is actually canceling (partly or completely, depending on its phase) a larger non-zero value of $\tilde{\kappa}_{e-}^{ZZ}$, so one can not set a valid upper limit simply equal to the measured value. One way to set a bound amongst the systematic is to include more than one independent set of data, (i.e. $n >1$ where $n$ is the number of data sets). A bound can then be set  by treating the amplitude of $\tilde{\kappa}_{e-}^{ZZ}$ as a statistic. This is possible because the phase of the systematic depends on the initial experimental conditions (i.e. phase with respect to the frame of reference of the test), and is likely to be random across the $n$ data sets \cite{prl,bchap}. If we take the mean of the $n$ $\tilde{\kappa}_{e-}^{ZZ}$ amplitudes, the systematic signal will cancel if the phase is random, but the possible Lorentz violating signal will not. Thus a limit can be set by taking the mean and standard deviation of the amplitude over the $n$ data sets. Therefore, unless more than one data set is analyzed a bound on the value of $\tilde{\kappa}_{e-}^{ZZ}$ can not be given in the presence of an unknown systematic. Since Antonini et. al. \cite{Ant} only gave statistics for one data set of 76 hours duration $(n=1)$, it is not possible to quote a bound on $\tilde{\kappa}_{e-}^{ZZ}$ from the analysis of this data.

In our recent rotating experiment we determined a value of $\tilde{\kappa}_{e-}^{ZZ}$ of $4.1(0.5)\times10^{-15}$ by fitting the amplitude over 5 data sets \cite{prl,bchap}. However, we did not use this result to claim an upper limit on the value of $\tilde{\kappa}_{e-}^{ZZ}$. Instead we followed the approach suggested above and took the mean and standard deviation of the amplitudes obtained from the individual data sets. This allowed us to determine $\tilde{\kappa}_{e-}^{ZZ}= 2.1(5.7)\times 10^{-14}$, the uncertainty being dominated by the variation of the systematic over the data sets. 

Antonini et. al. also claim a ten-fold improvement in the RMS \cite{Robertson,MaS} parameter $\beta-\delta-1/2$ \cite{Ant}, which is determined to be $(+0.5\pm 3\pm 0.7)\times 10^{-10}$, in comparison to the previous best result of $(-2.2\pm 1.5)\times 10^{-9}$ \cite{Muller}. Comparing the uncertainties this is no more than a factor of 5 improvement and has been overstated by a factor of 2. Our recent concurrent work obtained a value of $(+0.9\pm 2)\times 10^{-10}$, which is a factor of 7.5 improvement \cite{prl}.

\begin{acknowledgments}
This work was funded by the Australian Research Council.
\end{acknowledgments}


\begin{thebibliography}{99}
\bibitem{Lipa} Lipa J.A. et al. Phys. Rev. Lett. {\bf 90}, 6, 060403, (2003).
\bibitem{Muller} M\"uller H. et al., Phys. Rev. Lett. {\bf 91}, 2, 020401, (2003).
\bibitem{WolfGRG} Wolf P., et al., Gen. Rel. and Grav., {\bf 36}, 10, 2351, (2004).
\bibitem{Wolf04} Wolf P, et al., Phys. Rev. {\bf D70}, 051902(R), (2004).
\bibitem{Kosto1} Colladay D., Kosteleck\'y V.A., Phys.Rev.{\bf D55}, 6760, (1997); Colladay D., Kosteleck\'y V.A., Phys.Rev.{\bf D58}, 116002, (1998)
\bibitem{KM} Kosteleck\'y V.A.,Mewes M.,Phys.Rev.{\bf D66},056005,(2002).
\bibitem{Ant} Antonini, M. Okhapkin, E. G\"{o}kl\"u and S. Schiller, Phys. Rev. A {\bf 71}, 050101R 2005.
\bibitem{prl} Stanwix P.L., Tobar, M.E., Wolf P. et al., accepted Phys. Rev. Lett. (2005), arXiv: hep-ph/0506074.
\bibitem{bchap} Tobar M.E. Stanwix P.L., et. al., to be published in Springer Lecture Notes in Physics 2005, arXiv: hep-ph/0506200. 
\bibitem{Robertson} Robertson H.P., Rev. Mod. Phys. {\bf 21}, 378 (1949).
\bibitem{MaS} Mansouri R., Sexl R.U., Gen. Rel. Grav. {\bf 8}, 497, (1977).

\end{thebibliography}
\end{document}